\documentclass{article}
\usepackage{spconf,amsmath,graphicx}
\usepackage{tikz}
\usepackage{caption}
\usepackage{subcaption}
\usepackage{pgfplots}
 
\newcommand{\x}{\mathbf{x}}

\newcommand{\comment}[1]{}

\usepackage{cite}
\usepackage{graphicx}
\usepackage{float}
\usepackage{url}
\usepackage{amsmath,amsfonts,amssymb}
\usepackage{booktabs}
\usepackage{subcaption}
\usepackage[export]{adjustbox}
\usepackage[update,prepend]{epstopdf}
\usepackage[lined,algonl,ruled]{algorithm2e}
\usepackage{balance}
\usepackage{acronym}
\usepackage{color}
\usepackage{multirow}
\usepackage{multicol}

\newcommand{\argmax}{\operatornamewithlimits{argmax}}

\newcommand{\rrho}{\boldsymbol{\rho}}
\newcommand{\ttheta}{\boldsymbol{\theta}}

\newcommand{\vv}{\mathbf{v}}

\acrodef{DSE}{deep single expert}
\acrodef{AIR}{acoustic impulse response}
\acrodef{PSD}{power spectral density}
\acrodef{DNN}{deep neural network}
\acrodef{MFCC}{mel-frequency cepstral coefficients}
\acrodef{MMSE}{minimum mean square error}
\acrodef{ASR}{automatic speech recognition}
\acrodef{STSA}{short-time spectral amplitude estimator}
\acrodef{LSAE}{log-spectral amplitude estimator}
\acrodef{OMLSA}{optimally modified log spectral amplitude}
\acrodef{IMCRA}{improved minima controlled recursive averaging}
\acrodef{STFT}{short-time Fourier transform}
\acrodef{DFT}{discrete Fourier transform}
\acrodef{MoG}{mixture of Gaussians}
\acrodef{MoE}{mixture of experts}
\acrodef{ATF}{acoustic transfer function}
\acrodef{MoDE}{mixture of deep experts}
\acrodef{DMoE}{deep mixture of experts}
\acrodef{r.v.}{random variable}
\acrodef{p.d.f.}{probability density function}
\acrodef{NN}{neural network}
\acrodef{EM}{expectation-maximization}
\acrodef{SPP}{speech presence probability}
\acrodef{CMVN}{cepstral mean and variance normalization}
\acrodef{NN-MM}{neural network mixture-maximum}
\acrodef{PESQ}{perceptual evaluation of speech quality}
\acrodef{SNR}{signal to noise ratio}
\acrodef{DAE}{deep auto-encoder}
\acrodef{LLR}{log likelihood ratio}
\acrodef{WSS}{weighted spectral slope}
\acrodef{Covl}{overall quality}
\acrodef{Csig}{speech distortion}
\acrodef{Cbak}{background distortion}
\acrodef{WSJ}{Wall Street Journal}
\acrodef{SVM}{support vector machine}
\acrodef{IBM}{ideal binary mask}
\acrodef{IRM}{ideal ratio mask}
\acrodef{ReLU}{rectified linear unit}
\acrodef{WER}{word error rate}
\acrodef{MM}{MixMax}
\acrodef{MOS}{mean opinion score}
\acrodef{MSE}{mean square error}
\acrodef{pDNN}{phoneme DNN}
\acrodef{cDNN}{classifier DNN}
\acrodef{gDNN}{gating DNN}
\acrodef{SGD}{stochastic gradient descent}
\acrodef{VAD}{voice activity detector}
\acrodef{STOI}{short-time objective intelligibility measure} 
\title{ Speech enhancement with mixture of deep experts with  clean clustering pre-training} 
\name{ Shlomo E. Chazan \qquad Jacob Goldberger \qquad Sharon Gannot }
\address{  Faculty of Engineering, Bar-Ilan University, Ramat-Gan, Israel
\\
\texttt{\{Shlomi.Chazan,Jacob.Goldberger,Sharon.Gannot\}@biu.ac.il}
}

\begin{document}
\ninept
\maketitle
	\begin{abstract}
		In this study we present a \acf{MoDE} neural-network architecture for single microphone speech enhancement. 
		Our architecture comprises a set of \acp{DNN}, each  of which is an `expert' in a different  speech spectral pattern such as phoneme. A gating  \ac{DNN} is responsible for  the latent variables which are the weights assigned to each expert's output given a speech segment.  The experts estimate a mask from the noisy input and the final mask is then obtained  as a weighted average of the experts'  estimates, with the weights determined by the gating \ac{DNN}.  A soft spectral attenuation, based on the estimated mask, is then applied to enhance the noisy speech signal. As a byproduct, we gain reduction at the complexity in test time. We show that the experts specialization allows better robustness to unfamiliar noise types.\footnote{This project has received funding from the European Union’s Horizon 2020
Research and Innovation Programme under Grant Agreement No. 871245 and was supported by the Ministry of Science \& Technology,
Israel.}

	\end{abstract}
	

	

	\section{Introduction}
	
	A plethora of  approaches to solve the problem of speech enhancement using a single microphone can be found in the literature (see e.g.~\cite{38}). 
Although microphone array algorithms are nowadays widely used, there are still applications in which only a single microphone is available. However, the performance of current solutions is not always satisfactory.
		Classical model-based algorithms such as the \ac{OMLSA} estimator with the  \ac{IMCRA} noise estimator  were introduced to enhance speech signals contaminated by nonstationary noise signals~\cite{17,32}. Nevertheless, when the noisy input exhibit rapid changes in noise statistics, the processed signal tends to yield \emph{musical noise} artifacts at the output of the enhancement algorithm.
	
	In recent years, \ac{DNN}-based algorithms were derived to enhance noisy speech. A comprehensive summary of the common approaches can be found in \cite{wang2018supervised,unet_enhance2020yossi_adi}. Recent contributions in the field can be found in \cite{skipconvnet, single_mic_losses2020,lan2020combining}. 
		These  \ac{DNN}-based approaches have to cope with the large variability of  the speech signal. They are thus trained on huge databases with multiple noise types to cover the large variety of noisy conditions, especially in real-life scenarios~\cite{wang2015large}.
	
	To alleviate these obstacles, algorithms which take into account the variability of the speech were developed. In~\cite{das2012phoneme} and \cite{chazan2016hybrid}, the phoneme labels were used to enhance each phoneme separately. Yet, the capabilities of the DNN were only partly utilized. Phoneme-based architecture was introduced for \ac{ASR} applications~\cite{wang2016phoneme}. In this architecture, a set of \acp{DNN} was separately trained with an individual database, one for each phoneme,  to find the \ac{IRM}. Given a new noisy input, the \ac{ASR} system outputs the index of the phoneme associated with the current input, and that phoneme \ac{DNN} is activated to estimate the \ac{IRM}. This approach improved performance in terms of noise reduction and more accurate \ac{IRM} estimation. However,  when the \ac{ASR} system is incorrect, a wrong \ac{DNN} is activated. Additionally, the continuity of the speech is disrupted by mistakes in the \ac{ASR} system. Finally, the \ac{ASR} was not learned as part of the training phase.

In this work, we present  a \ac{MoDE} modeling  for speech enhancement.
The noisy speech signal comprises several different subspaces which have different relationships between the input and the output.
Each expert is responsible for enhancing  a single speech subspace and the  gating network finds the suitable weights for each subspace in each time frame. Each expert estimates a mask and  the  local mask decisions are averaged, based on the gating network, to provide the final mask result.
Since the gate is trained to assign an input to one of the experts in an unsupervised manner, random initialization of the MoDE parameters may be insufficient, as it tends to utilize only few of the experts. A pretraining stage, comprised of a clustering of clean speech utterances, is therefore applied in order to capture the speech variability and to alleviate this degeneration problem. {The clustering labels of the clean dataset are utilized for pre-training all experts and the gate network as well.}

The contribution of this work is twofold. First, we present a \acf{MoDE}-based enhancement procedure that automatically decomposes the speech space into simpler subspaces and applies a suitable different enhancement procedure for each input subspace. 	
  Second, we propose an  algorithm to train the  MoDE model that  does not require  a phoneme-labeled database.

	\section{Problem formulation }\label{sec:problem}
	
	Let $x(t)=s(t)+n(t)$ denote the observed noisy signal at discrete-time $t$, where $s(t)$ denotes the clean speech signal and $n(t)$ an additive noise signal.
		The \ac{STFT}  of $x(t)$  with frame-length $L$ is denoted by $\bar{x}_k(n)$, where $n$ is the frame-index and $k=0,1,\ldots,L-1$ denotes the frequency index. Similarly, $\bar{s}_k(n)$ and $\bar{n}_k(n)$ denote the \ac{STFT} of the speech and the noise-only signals, respectively.

    Different speech activation masks were proposed \cite{targetsDNN_IBMIRM,wang2018supervised}, and the most commonly used mask is the \acf{IRM}. The IRM of a single frame is defined as follows:
    \begin{equation}
    \text{IRM}_k=\left( \frac{|\bar{s}_k|^2}{|\bar{s}_k|^2+|\bar{n}_k|^2} \right) ^\gamma,
    \label{IRM}
    \end{equation}
    where  $\gamma$ is commonly set to  $\gamma=0.5$.
	
    We can cast the speech enhancement problem as finding an estimate $\rho_{k} \in [0,1]$ of the IRM mask $\text{IRM}_k$ by only using noisy speech utterances. The DNN task, is therefore to find the mask $\rho_k$, given the noisy signal.

 	In the enhancement task, only the noisy signal $\bar{x}$ is observed, and the goal is to estimate $\hat{\bar{\mathbf{s}}}=\left[\hat{\bar{s}}_0,\ldots,\hat{\bar{s}}_{L/2} \right] $ of the clean speech $\bar{\mathbf{s}}=\left[\bar{s}_0,\ldots,\bar{s}_{L/2} \right] $.
    Once the estimated mask $ \rrho=\left[ {\rho_0,\ldots,\rho_{L/2}}\right] $  is computed,  the  enhanced signal  can be obtained by:
    \begin{equation}
    	\hat{\bar{\mathbf{s}}}=\bar{\mathbf{x}}  \odot\boldsymbol{\rho}
        \label{wang_enhance}
    \end{equation}
    where  $\odot$ is a element-wise  product (a.k.a. Hadamard product).
     	In this work, we use a softer version of  \eqref{wang_enhance} to enhance the speech signal:
	\begin{equation}
	\hat{\bar{\mathbf{s}}}= \bar{\x} \odot \exp\{-(1-\rrho)\cdot \beta\}.
	\label{final_soft_enh}
	\end{equation}
	  Note, that in frequency bins where $\rho_k=1$, namely where the clean speech is dominant, the estimated signal will be $\bar{s}_k=\bar{x}_k$.  However, using $\rho_k=0$ in \eqref{wang_enhance}, namely in noise-dominant bins, may result in \emph{musical noise} artifacts~\cite{30}\cite{31}. In contrast, using \eqref{final_soft_enh}, the attenuation in noise-dominant bins is limited to $\exp\{-\beta\}$, potentially alleviating the musical noise phenomenon.
	
	As input features to the IRM estimating network, we use the log-spectrum of the noisy signal at a single time frame, denoted by $\x=\log|\bar{\x}|$. The network goal is to estimate 
	the mask $\rrho$.

	\section{Deep Mixture of Experts for Speech Enhancement}\label{sec:dforen}

The \acf{MoE} model, introduced by Jacobs et al.~\cite{mixture_of_experts,jordan1994hierarchical},  provides important paradigms for inferring a classifier from data. 

\noindent{\bf Statistical model } We can view the \ac{MoE} model as a two step process that produces a decision $\rrho$ given an input feature set $\x$. We first use the gating function to select an expert and then  apply the expert to determine the output label. The index of the selected expert can be viewed as an intermediate  hidden random variable denoted by $z$.
Formally, the \ac{MoE} conditional distribution  can be written as follows:
\begin{equation}
p(\rrho|\x;\Theta) = \sum_{i=1}^m p(z=i|\x;\ttheta_g) p(\rrho|z=i,\x;\ttheta_i)
\label{likmoe}
\end{equation}
such that  $\x$  is the log-spectrum vector  of the noisy speech, $\rrho$ is the \ac{IRM} vector and $z$ is a speech spectral state; e.g., the phoneme identity or  any other indication of a specific spectral pattern of the speech frame.
The model parameter-set ${\Theta}$ comprises the parameters of the gating function, $\ttheta_g$, and the parameters $\ttheta_1,\dots,\ttheta_m$ of all $m$ experts.
We further assume that both the experts and the gating functions are implemented by \acp{DNN}, thus this model is dubbed \acf{MoDE}.

	The input to each expert \ac{DNN} is the noisy log-spectrum frame together with context frames. 
	All $m$ experts in the proposed algorithm are implemented by  \acp{DNN} with the same structure.
	The gating \ac{DNN} is fed with the corresponding \ac{MFCC} features denoted by $\vv$.
   MFCC, which  is based on frequency bands,  is a more compact representation than a linearly spaced log-spectrum  and  is known  for its better representation of sound classes~\cite{hermansky2013ASRproperties}. We  found that using the MFCC representation for the gating \ac{DNN}  both slightly improves performance and reduces the computational complexity.
The output layer that provides the mask decisions is composed of  $L/2+1$ sigmoid neurons, one for each frequency band. Let $\hat{\rrho}_i$ be the mask vector computed by the $i$-th expert.
The mask decision of the $i$-th expert and the $k$-th frequency bin is defined as:
	\begin{equation}
		\hat{\rho}_{i,k}=p(\rho_k|\x,z=i;\ttheta_i).
		\label{rho_ki}
	\end{equation}
\sloppy
\noindent{\bf Parameter inference } We next address the problem of learning the \ac{MoDE} parameters (i.e.~the parameters of the experts and the gating function) given a training dataset $\{(\x(1),\rrho(1)),\dots,(\x(N),\rrho(N))\}$, where $N$ is the size of the database.
   Our  loss function is following the training strategy proposed in \cite{mixture_of_experts}, which  prefers  error function that encourages expert specialization instead of cooperation:
   \begin{multline}
   L(\Theta) =  - \sum_{n=1}^N \log \left( \sum_{i=1}^m p_i(n)  \exp( -d(\rrho(n),\hat{\rrho}_i(n))) \right)
\label{loss}
\end{multline}
such that $$p_i(n)=p(z(n)=i|\vv;\ttheta_g)$$ is the gating soft decision and $$\hat{\rrho}_i(n)=p(\rrho(n)|z(n)=i,\x(n);\ttheta_i)$$ is the $i$-th network prediction.
 We set $d(\rrho(n), \hat{\rrho}_i(n))$   to be the \ac{MSE} function between $\rrho(n)$ and $\hat{\rrho}_i(n)$,  i.e. $d(\rrho(n), \hat{\rrho}_i(n))=\frac{1}{2}\|\rrho(n)- \hat{\rrho}_i(n)\|^2$.

    To train the network parameters we can apply the standard back-propagation techniques.
     The gradients of the \ac{MoDE} parameters provide another intuitive perspective on the model.
    It can be easily verified that the back-propagation equations for the parameter set  of the $i$-th expert are:
\begin{equation}
\frac{\partial L}{\partial {\ttheta}_{i}} = \sum_{n=1}^N {w}_i(n)  (\rrho(n)-\hat{\rrho}_i(n))\cdot \frac{\partial} {\partial \ttheta_{i}}
   \hat{\rrho}_i(n)
\label{fder1}
\end{equation}
such that ${w}_i(n)$ is   the
posterior probability
of expert $i$:
\begin{equation}
    {w}_i(n)= p(z(n)=i|\x(n),\rrho(n);\Theta). 
\label{sestep}
\end{equation}

{Note, that $p_i(n)$ is the posterior probability of expert $i$
given the MFCC, and ${w}_i(n)$ is the posterior given the true label and the input.
}
Similarly, the  back-propagation equation for the parameter set of the gating \ac{DNN}  is:
\begin{equation}
\frac{\partial L}{\partial {\ttheta}_{g}} = \sum_{t=1}^N  \sum_{i=1}^m {w}_i(n) \cdot \frac{\partial }{\partial \ttheta_g} \log  p_i(n).
\label{fder2}
\end{equation}

During the training of the \ac{MoDE}, the gating DNN is learned  in an unsupervised manner. Namely, the input $\x$  propagates through all experts and the gate selects the output of one of the $m$ experts without any supervision. When dealing with a complex task  such as  clustering, parameter initialization is crucial.
In fact, without a smart initialization, trivial solution might occur and   only one or small number of experts will be activated by the gate. Therefore pretraining each expert as well as the gate DNNs is a must.

 In \cite{chazan2016iwaenc} the phoneme labels were first used  to train the gate as a phoneme classifier, and to train each expert with frames having the same phoneme. In our approach though, no labels are available.

In order to acquire labels in an unsupervised manner we propose to apply a clustering algorithm technique to the clean signals. The clustering is used to  find $m$ different patterns of the speech in the log-spectrum domain. The idea is that each cluster consists of frames with a similar acoustic pattern and therefore their masks are also expected to be similar. The clustering is applied to clean speech frames to encourage the clusters to focus on different speech patterns and not on different noise types.

We used clustering based on training of an autoencoder followed by a $k$-means clustering in the embedded space~\cite{Xie_2015}.   The obtained clustering  is used to initialize the MoDE parameters. The network components are then jointly trained using noisy speech data.

\noindent{\bf Network architecture} All $m$ experts in the proposed algorithm are implemented by  \acp{DNN} with the same structure. In addition to the current frame, the input features include four preceding frames and  four subsequent  frames to add context information; therefore, each input consists of nine frames.  The network consists of 3 fully-connected hidden layers with  512 \ac{ReLU} neurons each.
The output layer that provides the mask decisions is composed of  $L/2+1$ sigmoid neurons, one for each frequency band.

The architecture of the gating \ac{DNN}  is also composed of 3 fully connected hidden layers with 512 \ac{ReLU} neurons each.  The output layer here is a  softmax function that produces the gating distribution for the $m$ experts.

The log-spectrum of the noisy signal, $\x$, is only utilized as the input to the experts, and the gating \ac{DNN} is fed with the corresponding \ac{MFCC} features denoted by $\vv$ (also with context frames).

	\begin{table*}[thbp]
	\small
		\begin{center}
			\begin{tabular}{@{}lcc@{}}
				\toprule
				{\textbf{Train phase}}	& \textbf{Database} &  \textbf{Details} \\
				\midrule
				\textbf{DSE, S-MoDE, MoDE} & TIMIT (train set) &  white Gaussian , Speech-like, F-16 cockpit, restaurant   ,SNR=-5,5~dB\\
							\midrule
			{\textbf{Test phase}}	& \textbf{Database} &  \textbf{Details} \\
				\midrule
				\textbf{Speech} & TIMIT (test set)&\\
				\textbf{Noise} &	NOISEX-92  &  Room, Car, Babble, Factory   \\
				\textbf{{SNR}} &	- & -5, 0, 5, 10, 15~dB  \\
				\bottomrule
			\end{tabular}
		\end{center}
		\caption{Experimental setup.}
		\label{tabel:experiment_setup}
	\end{table*}

	\section{Experimental study}\label{sec:experiments}

\begin{figure}[t!!!]
	\begin{subfigure}[b]{0.25\textwidth}
		\centering
		\includegraphics[trim=0 0 0 0 ,clip ,  width=\textwidth]{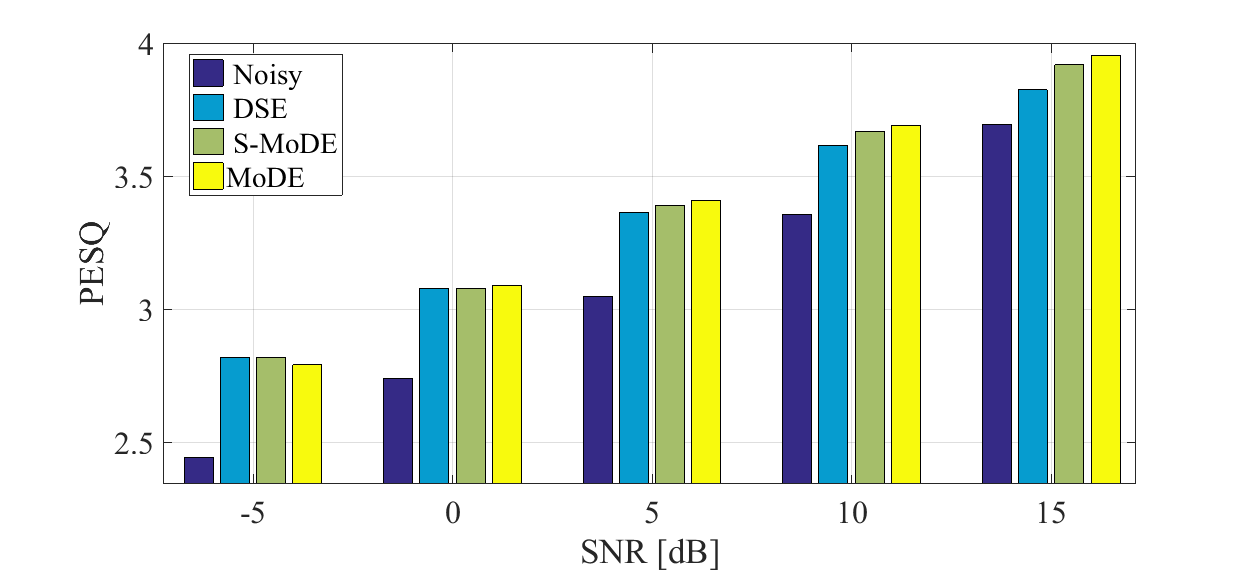}
		\caption{Car noise}
		\label{fig:pesq_car}
	\end{subfigure}
	\begin{subfigure}[b]{0.25\textwidth}
		\centering
		\includegraphics[trim=0 0 0 0 ,clip, width=\textwidth]{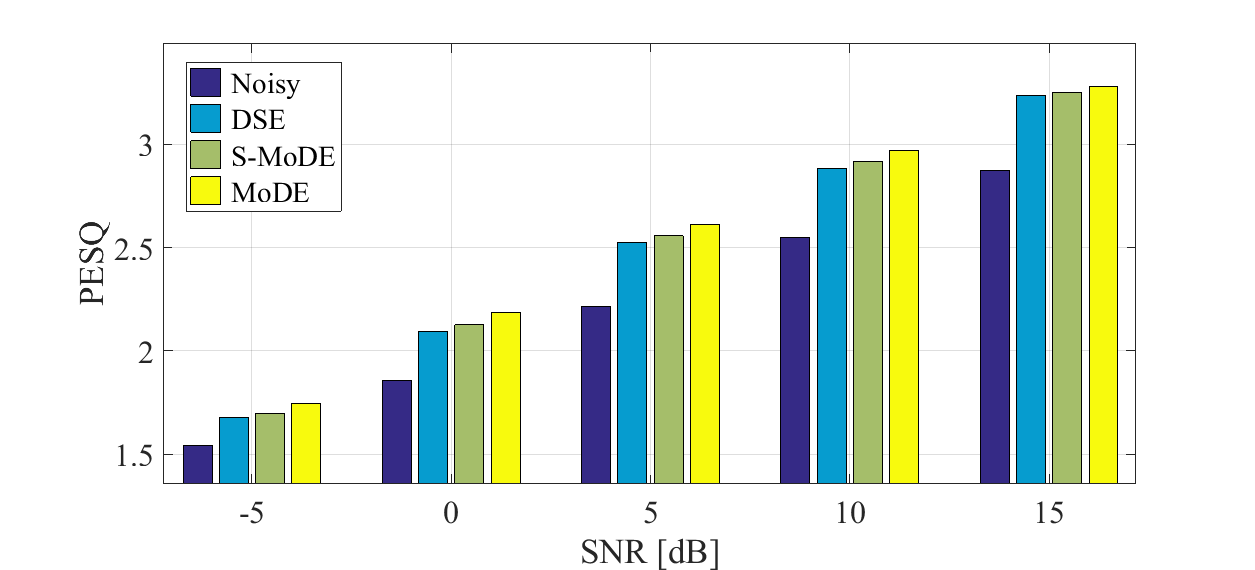}
		\caption{ Room noise}
		\label{fig:pesq_room}
	\end{subfigure}\\
	\begin{subfigure}[b]{0.25\textwidth}
		\hspace{0.1mm}
		\includegraphics[trim=0 0 0 0 ,clip,width=\textwidth]{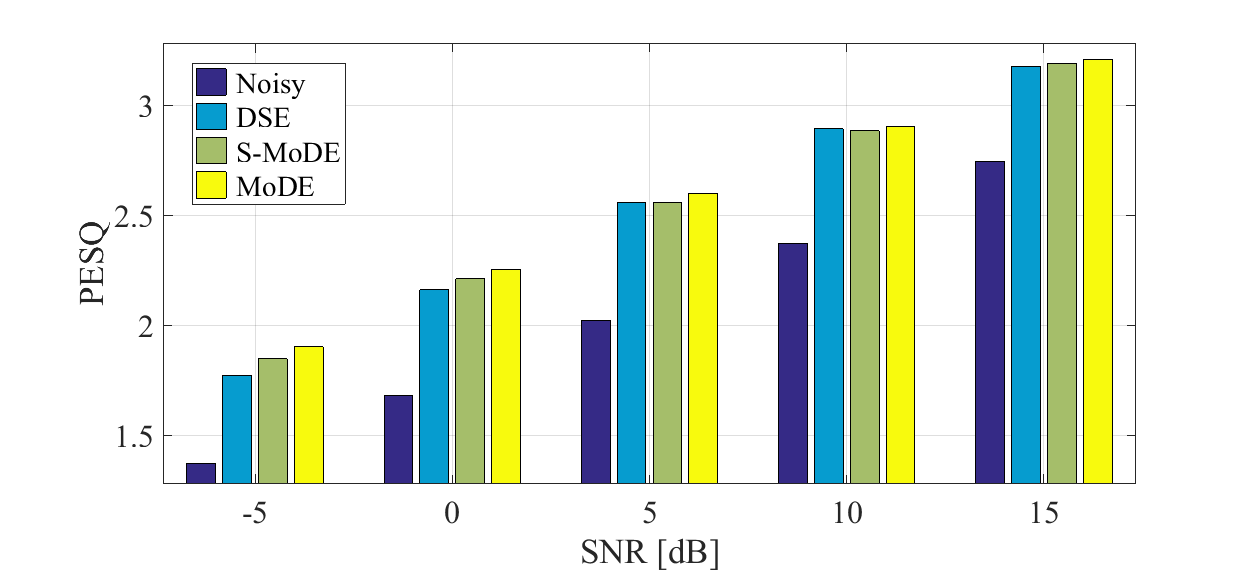}
		\caption{Factory noise}
		\label{fig:pesq_factory}
	\end{subfigure} %
	\begin{subfigure}[b]{0.25\textwidth}
		\includegraphics[trim=0 0 0 0 ,clip,width=\textwidth]{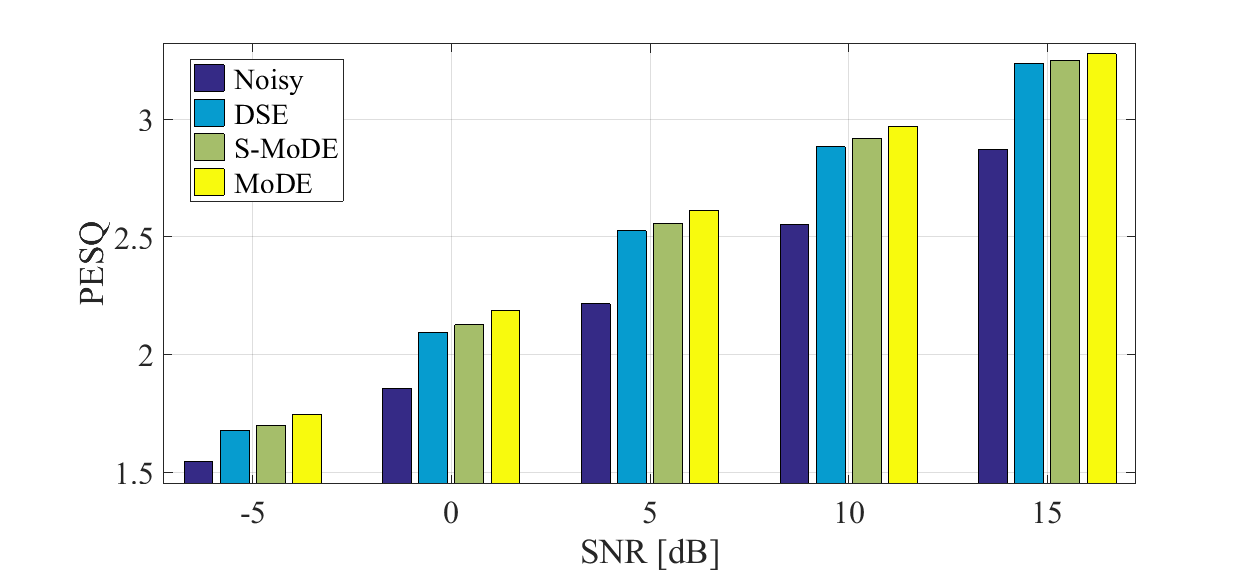}
		\caption{Babble noise}
		\label{fig:pesq_babble}
	\end{subfigure}
	\caption{PESQ results on various noise types. }
	\label{fig:pesq}
\end{figure}

	\begin{figure}[t!!!]
			\centering
			\begin{subfigure}[b]{0.25\textwidth}
				\includegraphics[width=\textwidth]{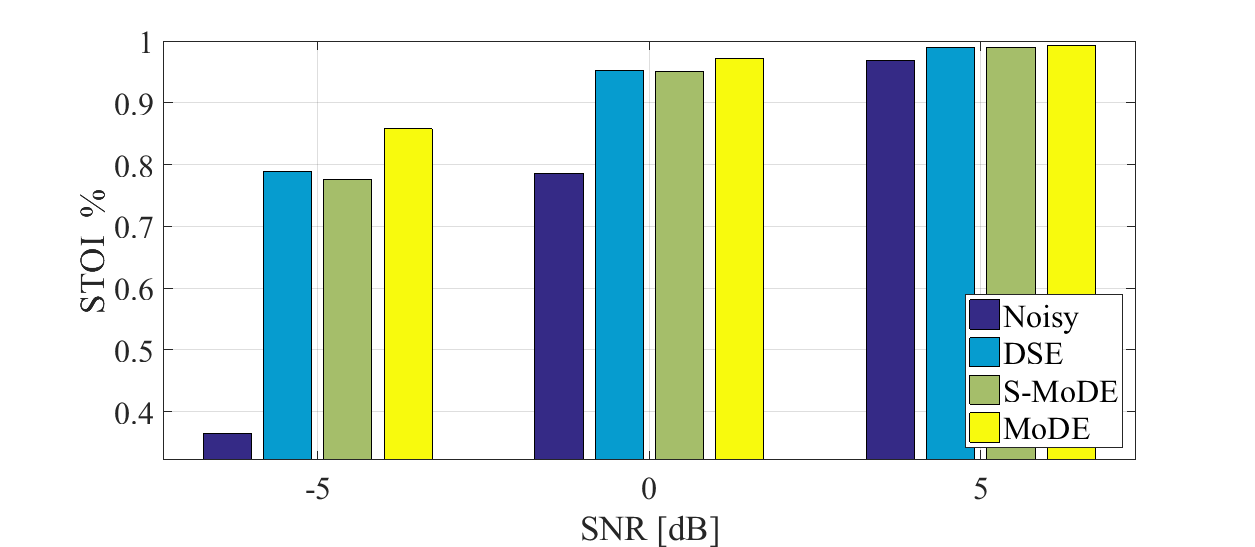}
				\caption{ {Speech} noise}
				\label{fig:stoi_Speech}
			\end{subfigure}%
			\begin{subfigure}[b]{0.25\textwidth}
				\includegraphics[width=\textwidth]{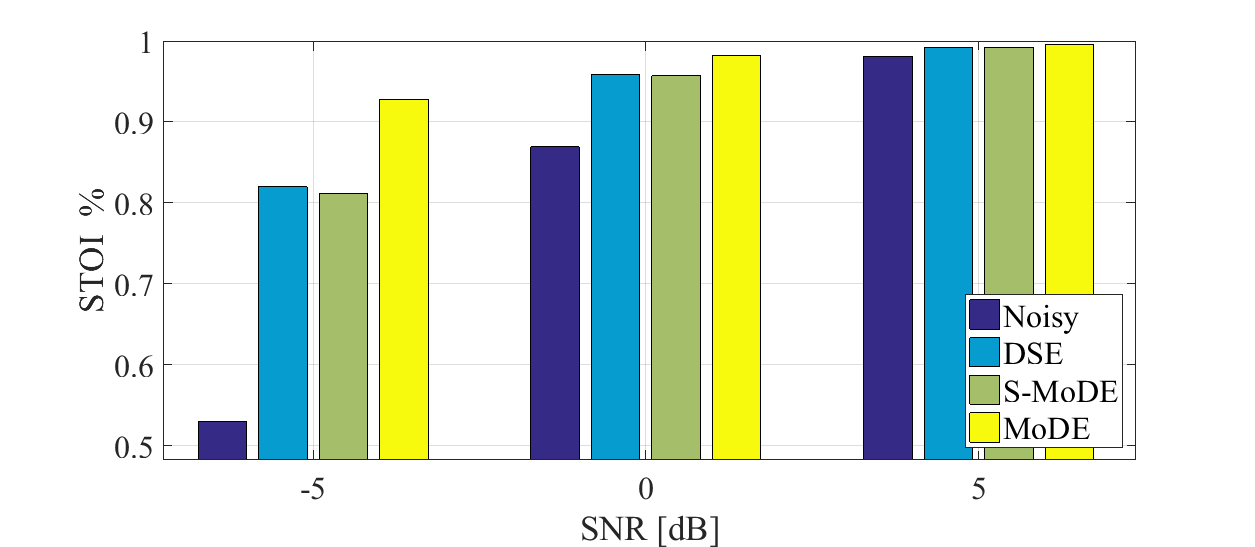}
				\caption{ {Room} noise}
				\label{fig:stoi_Room}
			\end{subfigure}\\		
			\begin{subfigure}[b]{0.25\textwidth}
				\includegraphics[width=\textwidth]{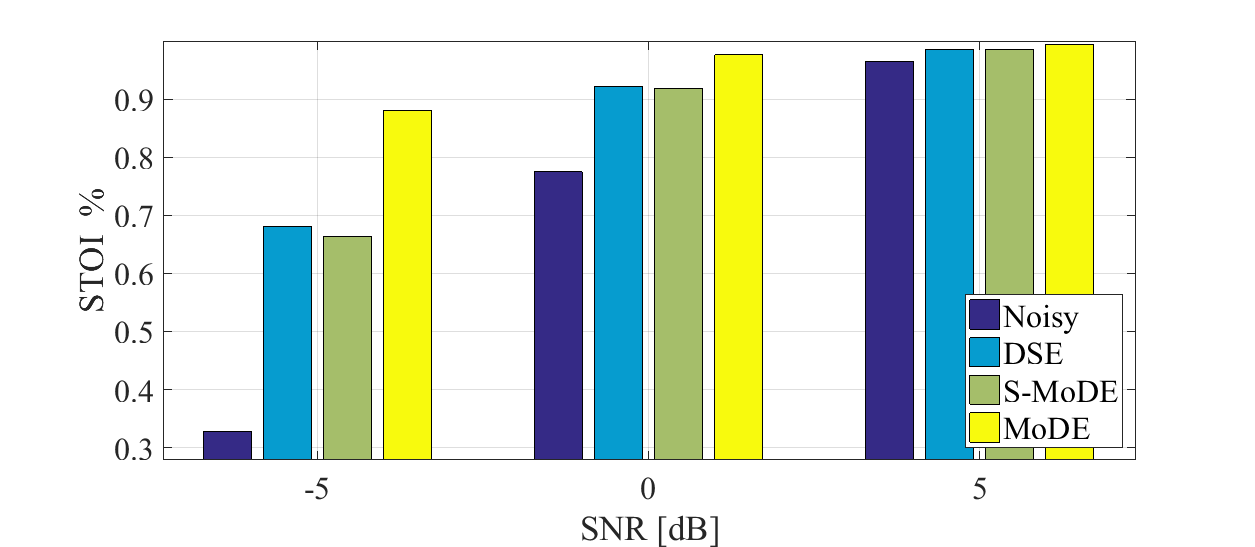}
				\caption{{Factory} noise}
				\label{fig:stoi_factory}
			\end{subfigure}%
			\begin{subfigure}[b]{0.25\textwidth}
				\includegraphics[width=\textwidth]{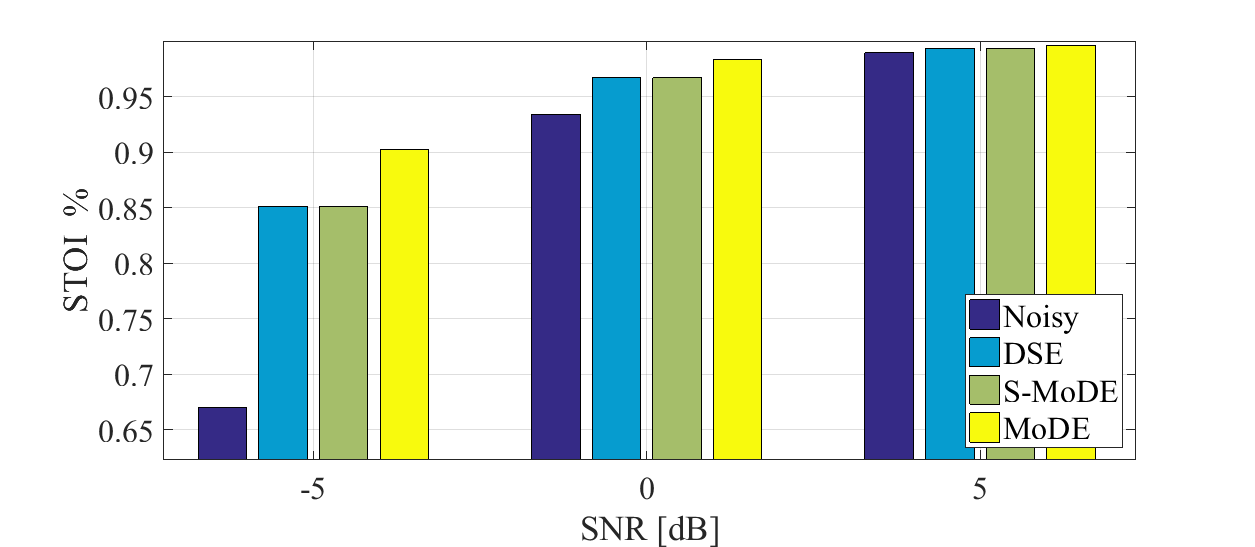}
				\caption{ {White} noise}
				\label{fig:stoi_white}
			\end{subfigure}%
			\caption{\ac{STOI} results for several noise types.}
			\label{fig:STOI}
		\end{figure}

	\noindent{\bf Setup}\label{subsec:expsetup}
	To test the proposed \ac{MoDE} algorithm we  contaminated the speech utterances with several types of noise from the NOISEX-92 database~\cite{43}, namely \emph{Car}, \emph{Room}, \emph{Factory} and \emph{Babble}.
	The noise was added to the clean signal drawn from the test set of the TIMIT database (24-speaker core test set), with  5 levels of \ac{SNR} at $-5$~dB, $0$~dB, $5$~dB, $10$~dB and $15$~dB chosen  to represent various real-life scenarios.

	\noindent{\bf Compared methods}
	We compared the proposed \ac{MoDE} algorithm with two  \ac{DNN}-based algorithms:
	1) Deep single expert (DSE) is a fully-connected architecture that  can be viewed as a single-expert network; and 2)
	   S-\ac{MoDE}  is a \textit{supervised} phoneme-based \ac{MoDE} architecture~\cite{chazan2016iwaenc}.   The network has 39  experts where each expert is explicitly  associated with a specific phoneme and  training uses  the phoneme labeling available in the TIMIT dataset.

	When using the \ac{MoDE} algorithm we need to set the number of experts.
	In most  \ac{MoE} studies,  the number of experts   was determined by an exhaustive search~\cite{yuksel2012twenty}. We found that increasing the number of experts from one to five significantly improves the performance and that additional experts had little effect. Hence, we chose the simpler model and set $m=5$.
	Each expert component in  the  S-\ac{MoDE} network has the same network architecture as the expert block of the proposed  \ac{MoDE} model. The \ac{DSE} architecture is a single DNN chosen to have the same size (in terms of the total number of neurons in each hidden layer) of the MoDE with 5 experts, for fair comparison.

\noindent{\bf Training Procedure}\label{subsec:training_procedure}
All the compared \ac{DNN}-based algorithms were trained with the same database. We used the TIMIT database~\cite{TIMIT} \emph{train} set (contains 462 speakers with 3.14 hours) for the training phase and the test set (containing 168 speakers with 0.81 hours) for the test phase. Note, that the train and test sets of TIMIT do not overlap.
Clean utterances were contaminated by multiple noise types, stationary and non-stationary, with varying \acp{SNR}.
The speech diversity modeling provided by the expert set was found to be rich enough to handle noise types that were not presented in the training phase.

The inputs to all DNN-based algorithms are the log-spectra vector and its corresponding \ac{MFCC} vector. The log-spectra and the MFCC vectors were concatenated to form the input feature vector of the \ac{DSE} network.  In the case of \ac{MoDE}, log-spectra were used as the input of the expert network and MFCC coefficients were the input of the gating network.
  Additionally, all methods apply the same enhancement scheme using \eqref{final_soft_enh}, where we set $\beta$ to correspond to attenuation of $20$~dB, a value which yielded high noise suppression while maintaining low speech distortion.

 The network was implemented in tensorflow~\cite{abadi2016tensorflow} with ADAM optimizer~\cite{adam} and batch-normalization was applied   to each layer~\cite{batchnorm}. To overcome the mismatch between the training and the test conditions, each utterance was normalized prior to the training of the network, such that the sample-mean and sample-variance of the utterance were  zero and one, respectively~\cite{12}.
 In order to circumvent over-fitting of the \acp{DNN} to the training database, we first applied the \ac{CMVN} procedure to the input, prior to the training and test phases~\cite{12}.

\noindent{\bf Objective quality measure results}\label{subsec:objectmeasurs}
		To evaluate the performance of the proposed speech enhancement algorithm, the standard \ac{PESQ} measure, which is known to have a high correlation with subjective score~\cite{13}, was used.
Intelligibility improvement was also  evaluated using  \ac{STOI}  \cite{STOI}.
		We  also carried out informal listening tests with approximately thirty listeners.\footnote{Audio samples comparing the proposed \ac{MoDE} algorithm with the DSE  and the S-\ac{MoDE} can be found in \texttt{www.eng.biu.ac.il/gannot/speech-enhancement/\\speech-enhancement-using-a-deep-mixture-of-experts/}.}


	Figure \ref{fig:pesq} depicts the \ac{PESQ} results of all algorithms for the {Car}, {Room}, {Factory} and {Babble} noise types as a function of the input \ac{SNR}. Figure \ref{fig:STOI} depicts the \ac{STOI} results for the same experiment setup.
		It is evident that both \ac{MoDE} and S-\ac{MoDE}, which split the noisy speech enhancement task into simpler problems, outperform the fully-connected network \ac{DSE}.
	Moreover, the proposed method, \ac{MoDE}, even outperforms the supervised method, S-\ac{MoDE}, which exploits the phoneme information.
	This indicates that splitting the noisy data according to the phonemes is not an optimal strategy for enhancement and allowing the network to find by itself a suitable splitting of the data yields improved results.

\begin{figure}[t!!!]
\begin{subfigure}[b]{0.25\textwidth}
		\includegraphics[trim=0 30 0 20 ,clip,width=\textwidth]{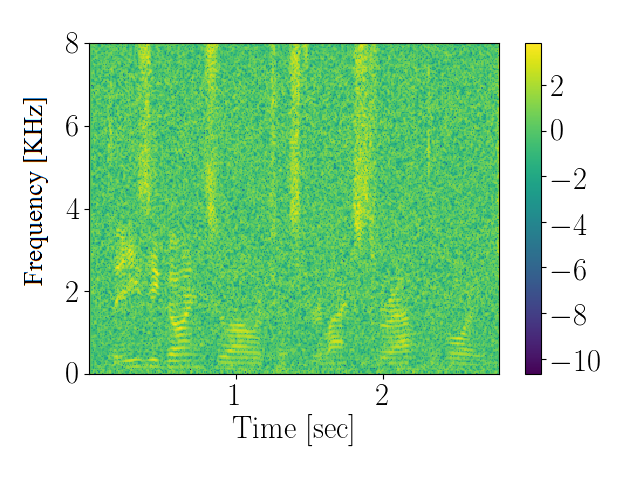}
		\caption{Noisy input.}
		\label{fig:noisy}
	\end{subfigure}
\begin{subfigure}[b]{0.25\textwidth}
		\includegraphics[trim=0 15 0 10 ,clip,width=\textwidth]{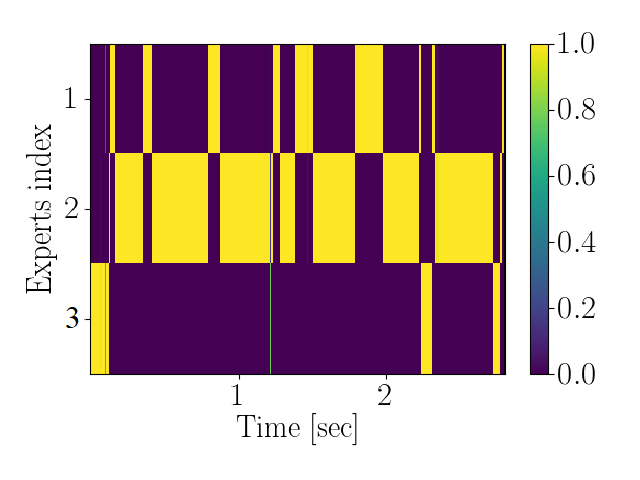}
		\caption{Gate output, $\hat{\mathbf{p}}=[\hat{p}_1, \hat{p}_2, \hat{p}_3]$.}
		\label{fig:gate_3}
	\end{subfigure}
	\begin{subfigure}[b]{0.25\textwidth}
		\hspace{0.1mm}
		\includegraphics[trim=0 30 0 20 ,clip,width=\textwidth]{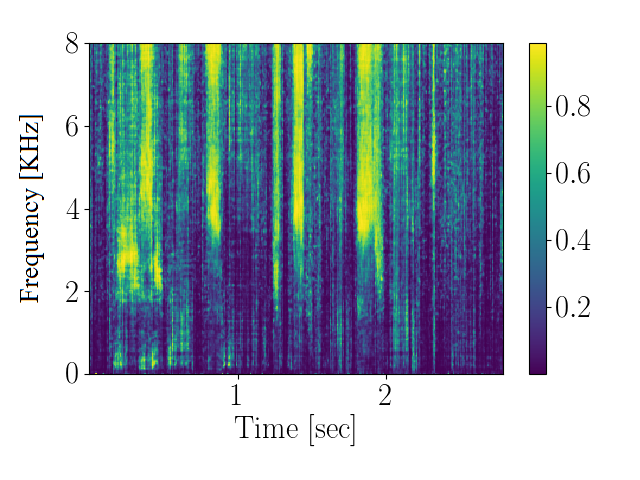}
		\caption{Expert 1 mask estimation, $\hat{\rrho}_1$.}
		\label{fig:ex_o_3}
	\end{subfigure}
	\begin{subfigure}[b]{0.25\textwidth}
		\includegraphics[trim=0 30 0 20 ,clip,width=\textwidth]{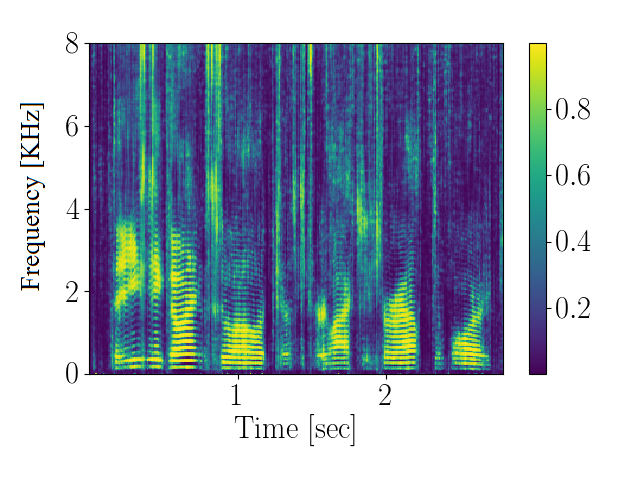}
		\caption{Expert 2 mask estimation, $\hat{\rrho}_2$.}
		\label{fig:ex_1_3}
	\end{subfigure}
	\begin{subfigure}[b]{0.25\textwidth}
		\includegraphics[trim=0 30 0 20 ,clip,width=\textwidth]{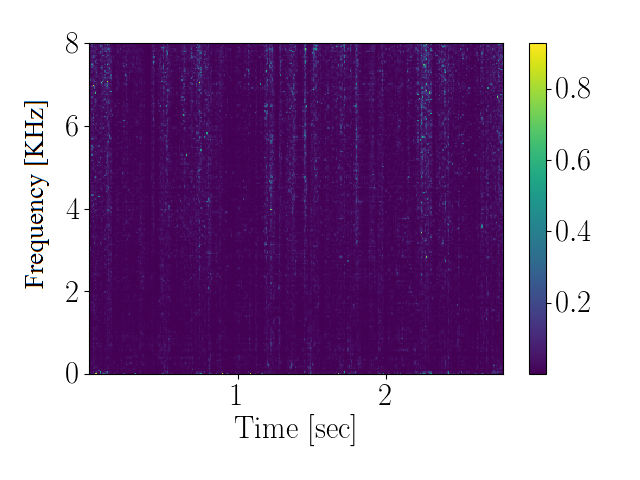}
		\caption{Expert 3 mask estimation, $\hat{\rrho}_3$.}
		\label{fig:ex_2_3}
	\end{subfigure}	\begin{subfigure}[b]{0.25\textwidth}
	\includegraphics[trim=0 30 0 20 ,clip,width=\textwidth]{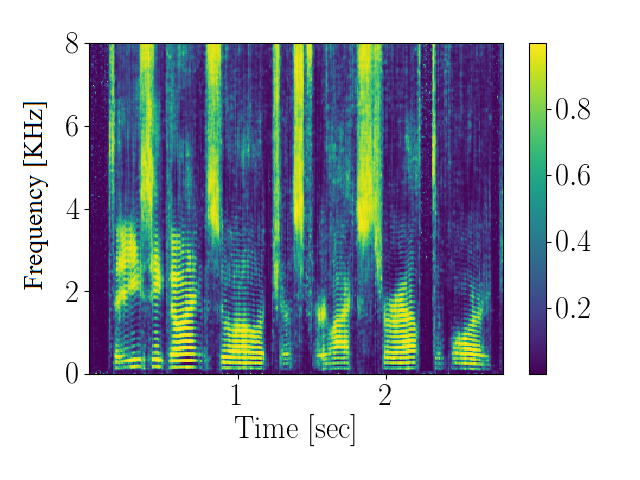}
	\caption{Final estimation, $\hat{\rrho}=\sum_{i=1}^3\hat{p}_i\cdot\hat{\rrho}_i$.}
	\label{fig:final_out_3}
    \end{subfigure}
	\caption{Experts and gate outputs of network with 3 experts. 
	}
	\label{fig:experties_mode3}
\end{figure}	

\noindent{\bf Interpretation of the learned experts}\label{subsec:experts}
	To gain a deeper understanding of the role of different  experts, we next present for each expert $i$ the mask estimation $\hat{\rho}_i$  for an example of noisy speech utterance contaminated by white noise SNR=5~dB (Fig.~\ref{fig:noisy}). Additionally, we show the distribution of the decisions of the gate DNN along the time.

	 In this case, the gate network classifies the noisy speech into  three classes, voiced frames, unvoiced frames and speech inactive frames (Fig.~\ref{fig:gate_3}). We can see that the  expertise of the first expert  is to enhance the unvoiced parts of the speech (Fig.~\ref{fig:ex_o_3}), while the second expert is  responsible for the voiced parts of the speech (Fig.~\ref{fig:ex_1_3}). Both experts do not perform well when the opposite speech pattern is introduced. The third expert expertise is to estimate the mask when only noise is present (Fig.~\ref{fig:ex_2_3}). The final weighted average masking decision is shown in Fig.~\ref{fig:final_out_3}.

	 We can also deduce from the gate decisions in 
	 Fig.~\ref{fig:gate_3}  that for each time-frame the gate DNN  tends to select only one expert. Consequently, each speech pattern is treated differently. Unlike the DSE DNN, in which a single network has to deal with the high variability of the speech patterns, our proposed method splits the speech enhancement task into $m$ simpler tasks, and  therefore outperforms the competing DSE.
	

This experiment suggests that each expert is responsible for a specific pattern of the speech spectrum. Consequently, the experts preserve the speech structure and a more robust behavior is exhibited compared to other \ac{DSE}-based algorithm.  
The S-MoDE do preserves the \emph{phoneme} structures with supervised learning. Yet, is seems that the unsupervised classification of the speech patterns is more beneficial.	

As a byproduct, we also gain complexity reduction at test time. For each time frame the gate first outputs a probabilities vector. The expert with the highest probability is therefore, $ i'=\argmax_i \{\hat{p}_i\} $.
Consequently, we can use only one expert for each time frame, 
\begin{equation}
    \hat{\rrho}=\sum_{i=1}^{m}\hat{p}_i \cdot \hat{\rrho}_i \approx \hat{\rrho}_{i'}. 
    \label{complexity}
\end{equation}
Therefore, even with larger number of experts, $m$, the same complexity of the gate and one of the expert is preserved. 


\section{Conclusion} \label{sec:summery}
This study introduced a \ac{MoDE} model for speech enhancement. This approach splits the  challenging task of speech enhancement into subspaces, where each \ac{DNN} expert is responsible for a simpler task which corresponds to a different speech type. The gating \ac{DNN} weights the outputs of the experts.
This approach makes it possible to  alleviate the well-known problem of \ac{DNN}-based algorithms, namely, the  mismatch between training phase and test phase. Additionally, the proposed \ac{MoDE} architecture enables training with a small database of noises and as a by product also reduce the complexity at test time.
The experiments verified that the proposed algorithm outperforms other \ac{DNN}-based approaches in both objective and subjective measures.


\balance
\bibliographystyle{IEEEbib}
\bibliography{ref}

\end{document}